\documentclass[a4paper]{jpconf}

\pdfoutput=1
    \usepackage[pdftex,     
            plainpages=false,   
            breaklinks=true,    
            colorlinks=true
           ]{hyperref} 
\usepackage{thumbpdf}
\usepackage{amsmath}
\usepackage{amssymb}
\usepackage{amsfonts}
\usepackage{array}
\usepackage{bbold}
\usepackage{cancel}
\usepackage[hang,small]{caption2}
\usepackage{subfigure}
\usepackage{verbatim}
\usepackage{slashed}
\usepackage{graphicx}
\usepackage{url}
\usepackage{enumitem}

\newcommand{\be}{\begin{equation}}
\newcommand{\ee}{\end{equation}}

\begin{document}

\title{The Game of Triangles}

\author{Michelangelo Preti }

\address{Nordita, KTH Royal Institute of Technology and Stockholm University, Roslagstullsbacken 23, SE-106 91 Stockholm, Sweden.}

\ead{michelangelo.preti@gmail.com}

\begin{abstract}
We present \texttt{STR} (Star-Triangle Relations), a Mathematica package designed to solve Feynman integrals by means of the method of uniqueness in any Euclidean spacetime dimension. We provide a set of tools to draw Feynman diagrams and interact with them only by the use of the mouse. Throughout the use of a graphic interface, the package should be easily accessible to users with little or no previous experience on diagrams computation.
\end{abstract}

\vspace*{-130mm}
\begin{flushright}
NORDITA 2019-047\\
$\;$\\
\footnotesize{
Contribution to the Proceedings of ACAT-2019\\ 
(Saas-Fee, Switzerland, March 10--15 2019)
}
\end{flushright}
\vspace*{105mm}

\section{Introduction}
\label{sec:intro}

Perturbative quantum field theories (QFTs) provide high-precision computations of various quantities in terms of Feynman diagrams. 
Since the number of diagrams grows rapidly with the perturbative order and the precision of numerical calculations is not often satisfactory, the development of analytic tools to compute multi-loop integrals plays a central role in QFT.
One of the most efficient technique is the \textit{integration by parts} method that allows to expand a diagram on a base of a finite number of \textit{master integrals} \cite{Vasiliev:1981yc}. However, this method alone is not always sufficient and it has to be combined with other techniques as the one of \textit{Gegenbauer polynomial} \cite{Chetyrkin:1980pr}, or \textit{Mellin transform} \cite{Bergere:1973fq,Usyukina:1975yg}, or \textit{HQET} (see \cite{Grozin:2014hna,Grozin:2015kna,Bianchi:2017svd} for recent applications in 4 and 3 dimensions), or \textit{differential equations} \cite{Remiddi:1997ny}. 

In this paper, we consider a powerful technique called \textit{method of uniqueness}, also known as \textit{star-triangle relation}.
This method provides a set of simple transformations to simplify Feynman integrals without performing any explicit integration. 
Introduced in \cite{DEramo:1971hnd}, it appears the first time in the context of conformal field theories in \cite{Fradkin:1978pp} and successively applied multi-loop calculations in \cite{Vasiliev:1981yc,Usyukina:1983gj,Kazakov:1983ns,Kazakov:1983pk,Kazakov:1984km}. Considering that such a sequence in general could be very long and in particular not unique, the development of a simple and automatic method is needed.

For these motivations, we present \texttt{STR} (\textbf{S}tar-\textbf{T}riangle \textbf{R}elations), a Mathematica package designed to solve Feynman integrals by means of the method of uniqueness for any Euclidean spacetime dimension $D$. The main feature of the package is to provide an user-friendly graphical interface in which the user can draw Feynman diagrams and interact with them only by the use of the mouse. The interactive window consists in some drawing tools to design the desired graph and a set of computational tools to act on the diagram through the star-triangle relations.
At any step of the process, the user can also print or export the output data (list of the uniqueness equations for the weights, the result of the computations and the integral representation of the diagram) and the graph itself. Any of these outputs is stored in a specific function that can be used in the current Mathematica session. For a more detailed manual, see \cite{Preti:2018vog} in which the package was originally presented.

\section{The method of uniqueness (star-triangle relations)}
\label{sec:startriangle}

We consider Feynman diagrams in coordinate space, where each vertex represents a point in the $D$-dimensional Euclidean space $\mathbb{R}^D$ while the lines with weights $\alpha_i$ are associated to the following massless propagators ($x_{ij}\equiv x_i-x_j$)
\begin{equation}\begin{split}\label{prop}
\vcenter{\hbox{\includegraphics[width=3.7cm]{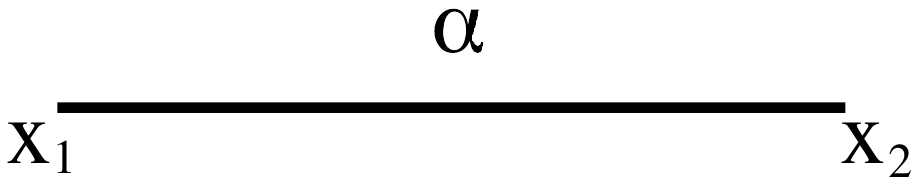}}}=\frac{1}{(x_{12}^2)^\alpha}\qquad\text{and}\qquad\vcenter{\hbox{\includegraphics[width=3.7cm]{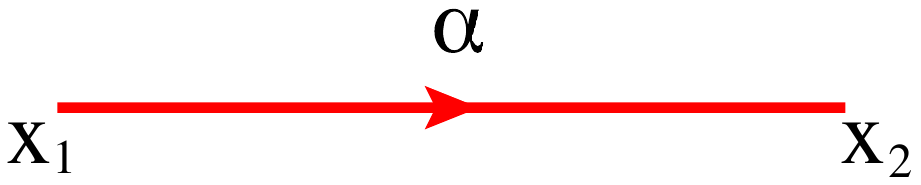}}}=\frac{\slashed{x}_{12}}{(x_{12}^2)^{\alpha+1/2}}\,,
\end{split}\end{equation}
respectively the scalar and spin-1/2 fermionic propagators where $\slashed{x}$ is the contraction between $x^\mu$ and the element of the Clifford algebra in $D$ dimensions. The momentum space propagators are defined by the following Fourier transforms
\begin{equation}\label{Fourier}
\frac{1}{(x_{12}^2)^\alpha}=\frac{\mathbb{a}_0(\alpha)}{4^\alpha\pi^{D/2}}\!\int \!\!d^Dk\frac{e^{ik\cdot x_{12}}}{(k^2)^{D/2-\alpha}}\quad\text{and}\quad
\frac{\slashed{x}_{12}}{(x_{12}^2)^{\alpha+1/2}}=\frac{-i\,\mathbb{a}_{1/2}(\alpha)}{4^\alpha\pi^{D/2}}\!\int\!\! d^Dk\frac{e^{ik\cdot x_{12}}\slashed{k}}{(k^2)^{D/2-\alpha+1/2}}\,,
\end{equation}
where
\begin{equation}\label{defa}
\mathbb{a}_{\ell}(\alpha)=\frac{\Gamma\left(\frac{D}{2}-\alpha+\ell\right)}{\Gamma(\alpha+\ell)}\qquad\text{with}\qquad\mathbb{a}_{\ell}(\alpha)\mathbb{a}_{\ell}(D/2-\alpha)=1\,.
\end{equation}
We call the \textit{weight of the diagram} (or of a portion of it) the sum of all the weights of the constituent lines. 
We shall say that a line, star and triangle\footnote{We define a star a three-leg vertex and a triangle a three-propagator loop.} are \textit{unique} if their weights are 0, $D$ and $D/2$ respectively. If a Feynman diagram contains unique stars or triangles, its computation is drastically simplified by means of the \textit{method of uniqueness} \cite{Kazakov:1983pk,Usyukina:1983gj,Belokurov:1983km,Kazakov:1984km,Kazakov:1984bw,Chicherin:2012yn}. 

This method consists in the set of rules listed below.
\begin{enumerate}
\item{\textbf{Merging rules:}} A set of identities to represent simple loop of propagators as a single line. 
\begin{itemize}
\item Simple loop of bosonic propagators
\begin{equation}\label{merge1}
\vcenter{\hbox{\includegraphics[width=2.9cm]{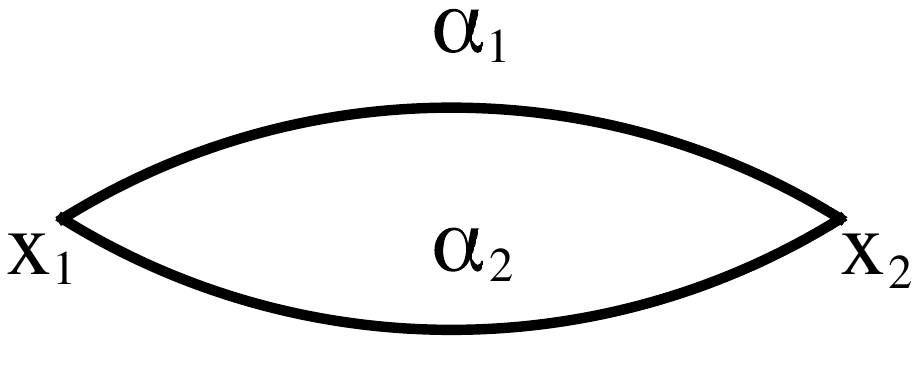}}}=\vcenter{\hbox{\includegraphics[width=2.7cm]{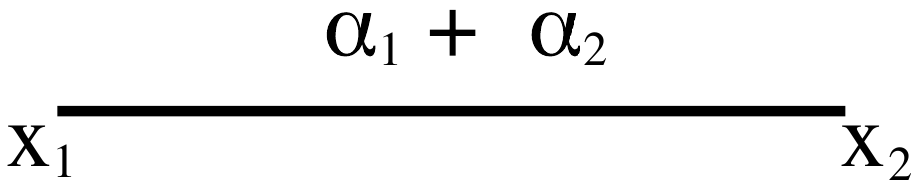}}}\,.
\end{equation}
\item Simple loop of mixed bosonic and a fermionic propagators
\begin{equation}\label{merge2}
\vcenter{\hbox{\includegraphics[width=2.9cm]{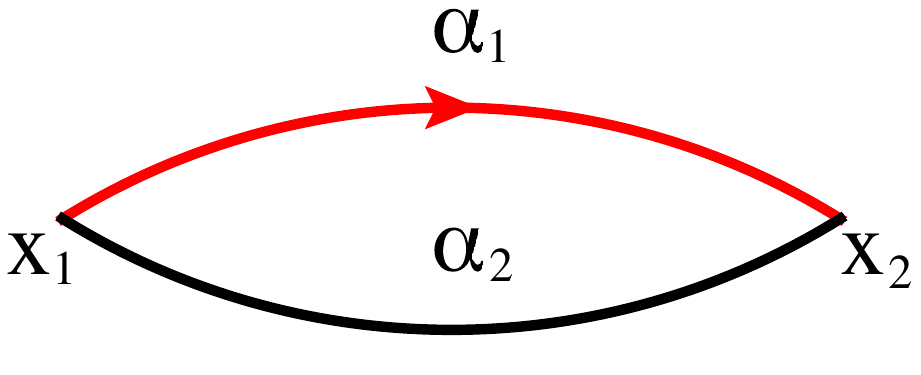}}}=\vcenter{\hbox{\includegraphics[width=2.7cm]{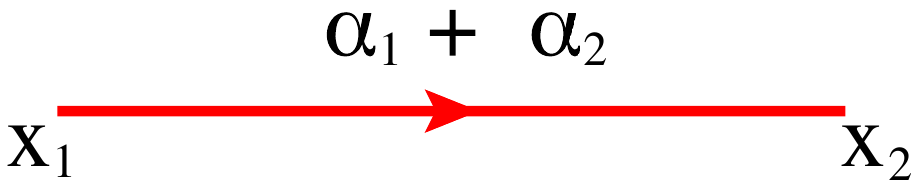}}}\,.
\end{equation}
\item Simple loop of fermionic propagators
\begin{equation}\label{merge3}
\vcenter{\hbox{\includegraphics[width=2.9cm]{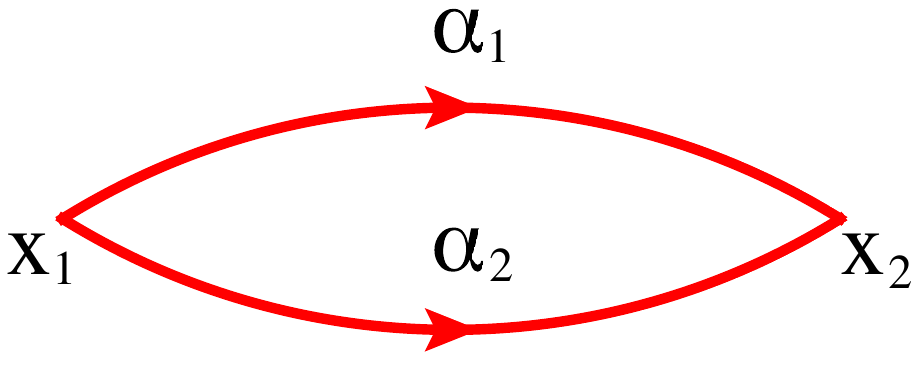}}}=\mathbb{1}\;\vcenter{\hbox{\includegraphics[width=2.7cm]{merge_bosbos2}}}\,.
\end{equation}
where we consider adjacent fermionic lines are contracted and the identity matrix $\mathbb{1}$ corresponds to $\mathbb{1}_{2^{D/2-1}}$ for even $D$ or $\mathbb{1}_{2^{(D-1)/2}}$ for odd $D$. 
\end{itemize}
\item{\textbf{Star-triangle relations:}} Those are the main identities of the uniqueness method that allow to integrate a unique star into a unique triangle and vice-versa. 
\begin{itemize}
\item The bosonic star-triangle relation 
\begin{equation}\label{STRbos}
\vcenter{\hbox{\includegraphics[width=2.8cm]{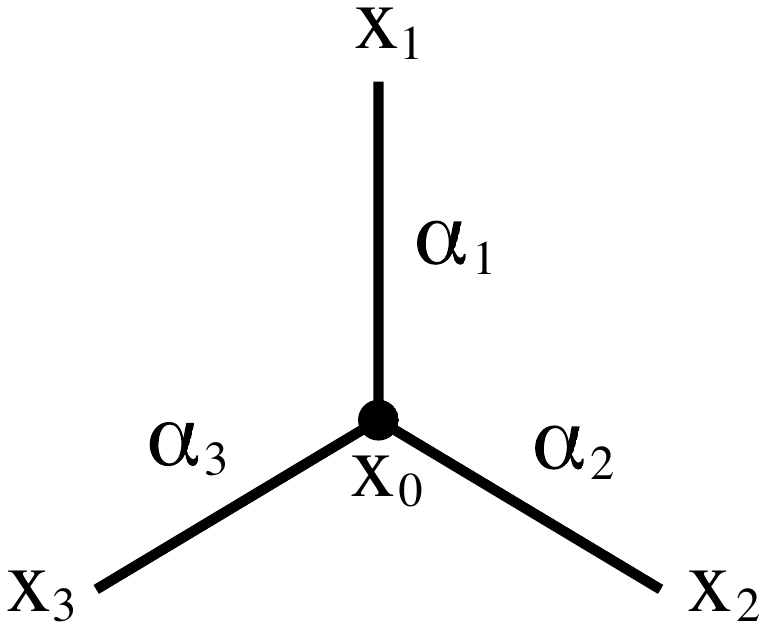}}}\overset{\sum_k\alpha_k=D}{=}
\quad\pi^{D/2}\;\mathbb{a}_{0}(\alpha_1,\alpha_2,\alpha_3)\;\vcenter{\hbox{\includegraphics[width=2.8cm]{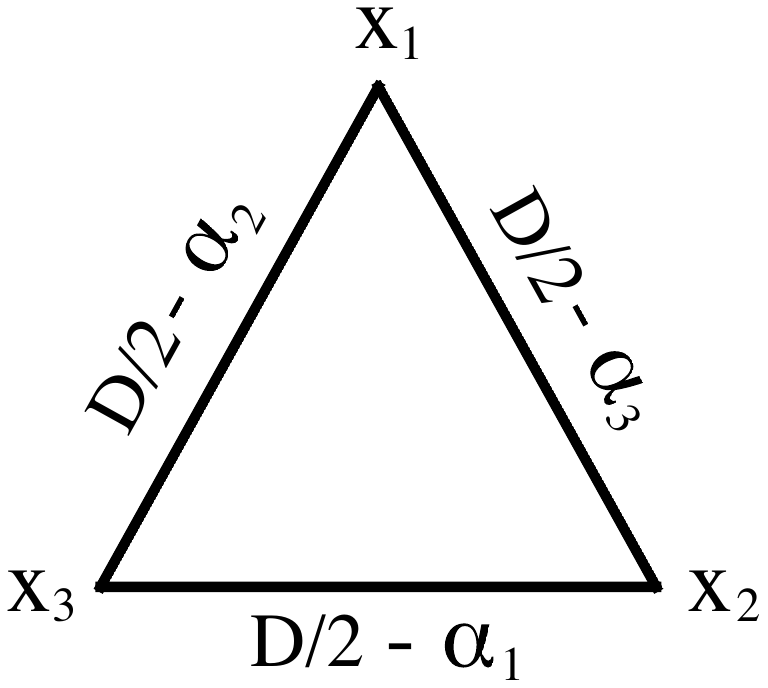}}}\,,
\end{equation}
where the function $\mathbb{a}_{\ell}$ with many arguments has the following property
\begin{equation}\label{amultiple}
\mathbb{a}_{\ell}(\alpha_1,\alpha_2,...,\alpha_n)=\prod_{k=1}^n \mathbb{a}_{\ell}(\alpha_k)\,.
\end{equation}
\item  The Yukawa star-triangle relation
\begin{equation}\label{STRferm}
\vcenter{\hbox{\includegraphics[width=2.8cm]{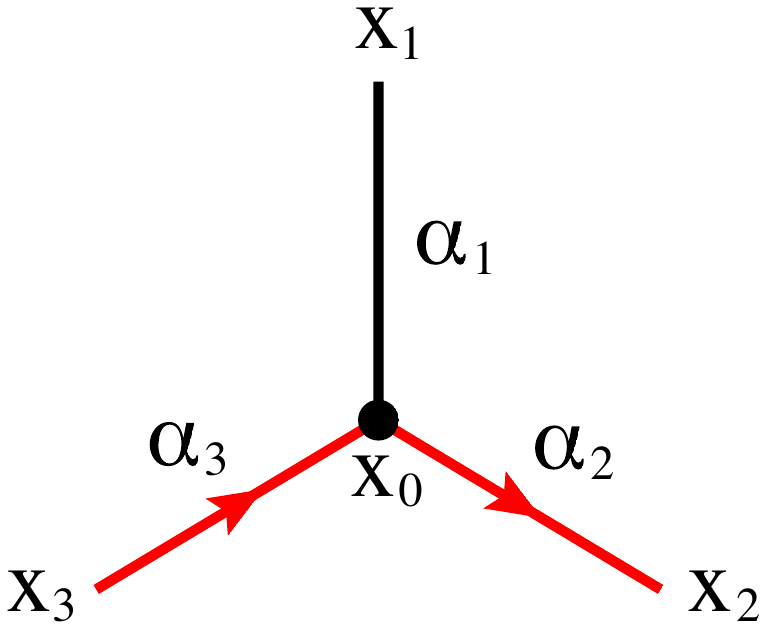}}}\overset{\sum_k\alpha_k=D}{=}
\quad\!\!\pi^{D/2}\;\mathbb{a}_{0}(\alpha_1)\;\mathbb{a}_{1/2}(\alpha_2,\alpha_3)\;\vcenter{\hbox{\includegraphics[width=2.8cm]{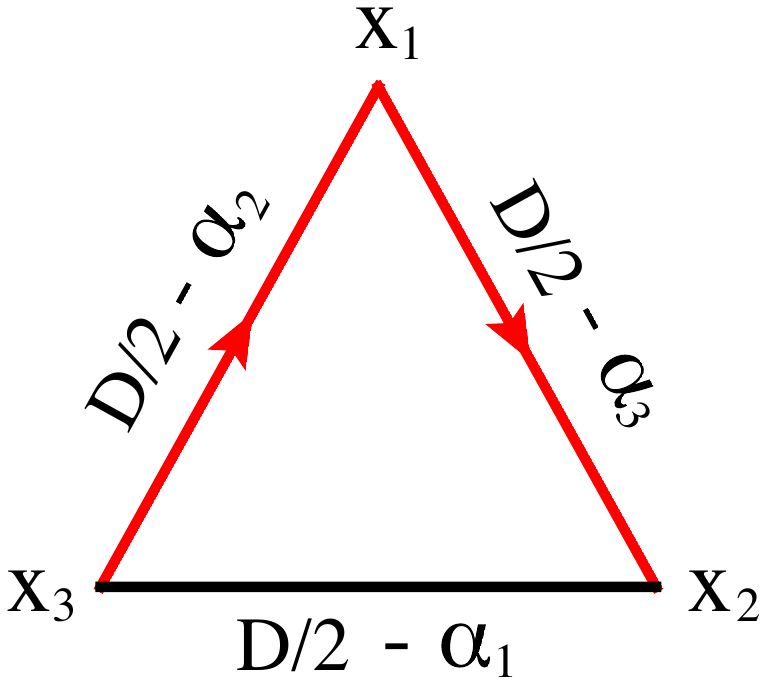}}}\,.
\end{equation}
\end{itemize}
\item{\textbf{Chain rules:}} A set of identities needed to integrate two propagators meeting in one internal (integrated) point (a simple loop in momentum space) in terms of a single propagator. 
\begin{itemize}
\item A chain of bosonic propagators
\begin{equation}\label{chain1}
\vcenter{\hbox{\includegraphics[width=3cm]{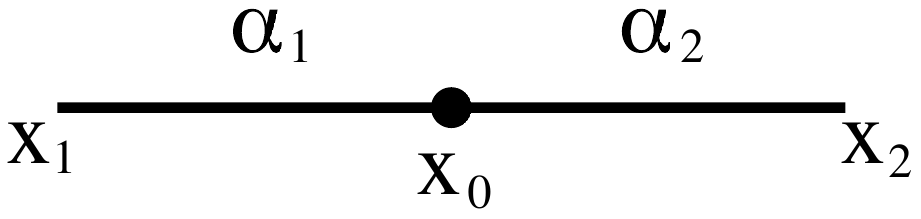}}}=\pi^{D/2}\;\mathbb{a}_{0}(\alpha_1,\alpha_2,D-\alpha_1-\alpha_2)\;\vcenter{\hbox{\includegraphics[width=3cm]{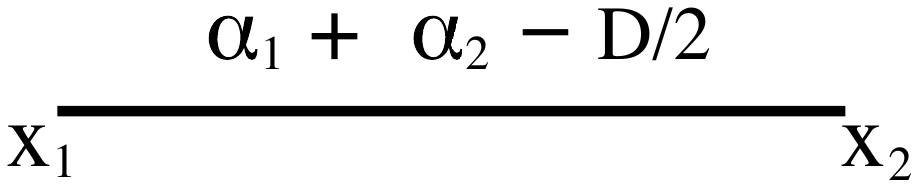}}}\,.
\end{equation}
\item A chain of mixed fermionic and bosonic propagators
\begin{equation}\label{chain2}
\vcenter{\hbox{\includegraphics[width=3cm]{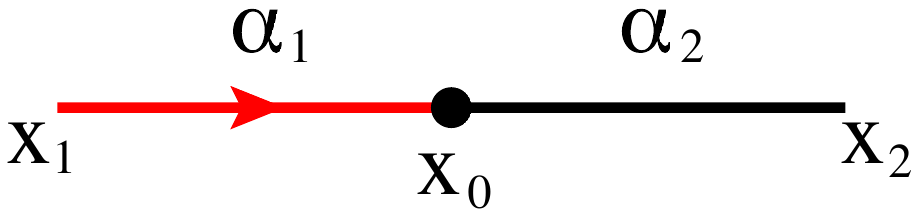}}}=\pi^{D/2}\;\mathbb{a}_{0}(\alpha_2)\,\mathbb{a}_{1/2}(\alpha_1,D-\alpha_1-\alpha_2)\;\vcenter{\hbox{\includegraphics[width=3cm]{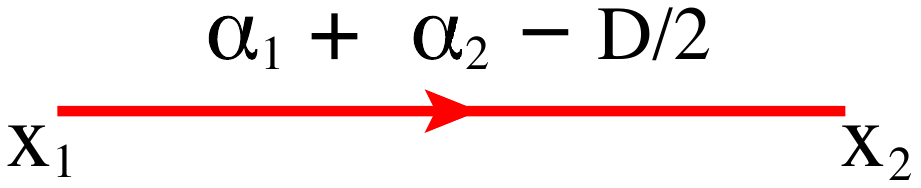}}}\,.
\end{equation}
\item A chain of fermionic propagators
\begin{equation}\label{chain3}
\!\!\vcenter{\hbox{\includegraphics[width=3cm]{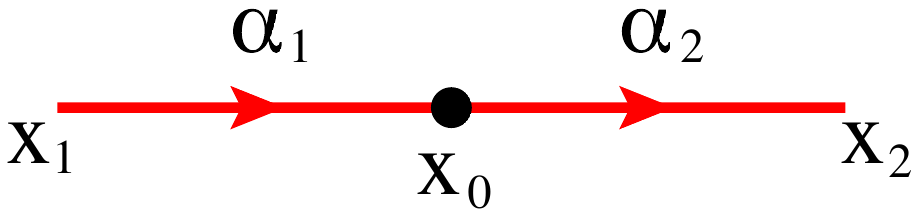}}}\!=\!-\pi^{D/2}\,\mathbb{a}_{0}(D-\alpha_1\!-\alpha_2)\,\mathbb{a}_{1/2}(\alpha_1,\alpha_2)\,\mathbb{1}\vcenter{\hbox{\includegraphics[width=3cm]{chain_bosbos2}}}\,.
\end{equation}
\end{itemize}
\end{enumerate}

\section{The \texttt{STR} package user guide}\label{sec:sec2}
The package can be downloaded from the GitHub repository at  \url{https://github.com/miciosca/STR}.
Once the package is loaded in a Mathematica session\footnote{For more detailed set up instructions see \cite{Preti:2017fjb,Preti:2018vog}}, the following
new functions will be available:
\begin{itemize}
  \item \texttt{STR[\textit{dimension}]}: specifying the dimension of the Euclidean spacetime \texttt{\textit{dimension}}, the function opens an interactive panel in which it is possible to draw and compute Feynman diagrams by means of the method of uniqueness (section \ref{sec:startriangle});
  \item \texttt{STRrelation}: a list of relations that identify unique stars and triangles in the diagram;
  \item \texttt{STRintegral}: the integral representation of the diagram drawn in \texttt{STR};
  \item \texttt{STRprefactor}: the result of acting on the graph with the uniqueness method;
  \item \texttt{STRgraph}: a modifiable version of the diagram drawn in \texttt{STR} as \texttt{Graphics[]} object;
    \item \texttt{STRSimplify[\textit{expr},\textit{dimension}]}: specifying the dimension of the spacetime \texttt{\textit{dimension}}, it rewrites \textit{expr} in terms of Euler gammas by means of \eqref{defa} and \eqref{amultiple}.
\end{itemize}
When the package is loaded, run the function \texttt{STR} specifying the dimension of the spacetime\footnote{The dimension can be a number, a letter or a combination of them with arithmetic operations.} without semicolon at the end.
\begin{figure}[t]
\begin{center}\label{STRpanel}
\includegraphics[width=0.85\textwidth]{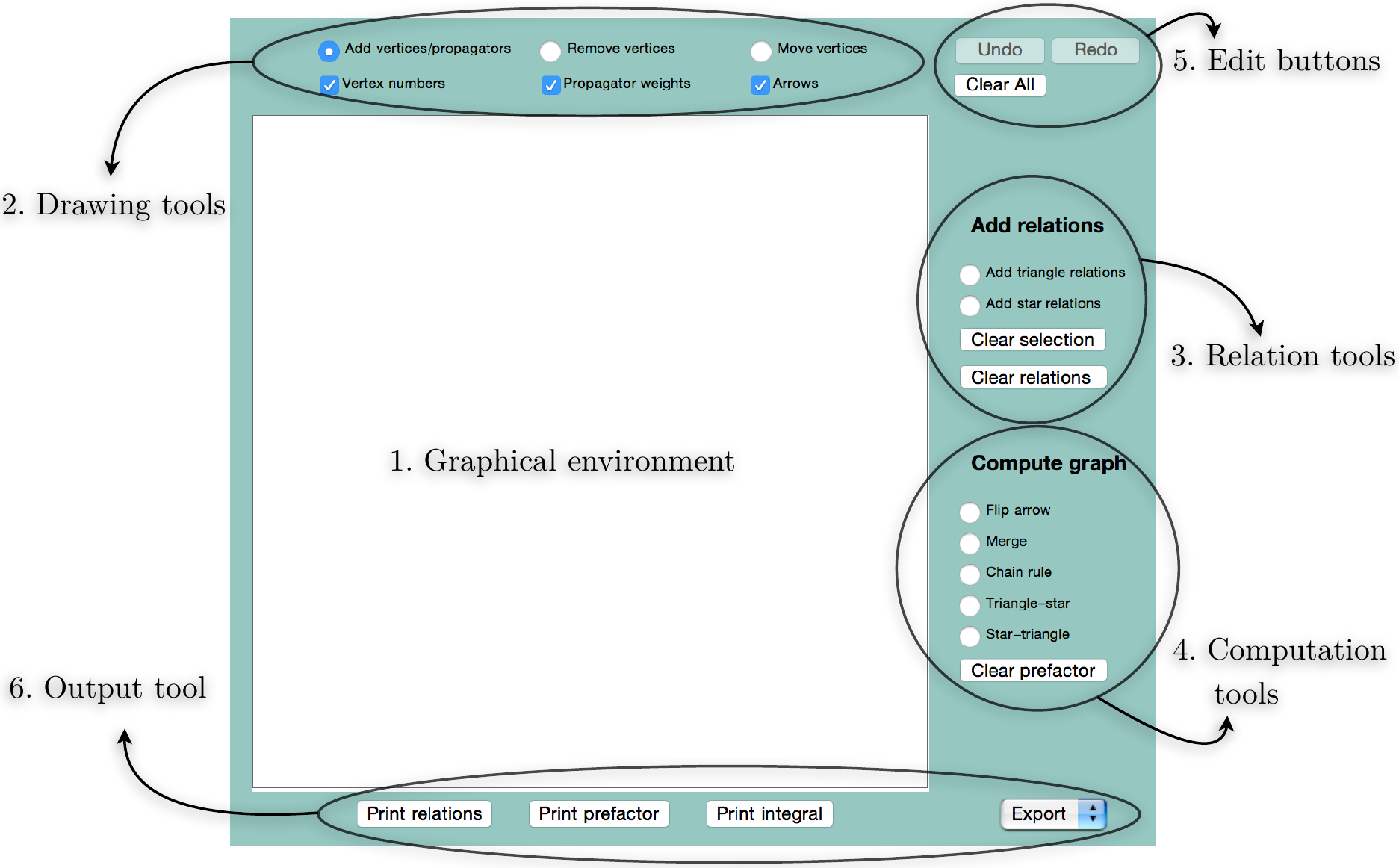}
\end{center}
\caption{The interactive window of the function \texttt{STR[\textit{dimension}]}.}
\label{fig:panel}
\end{figure}
The output will be the interactive panel in Figure \ref{fig:panel}. The user can draw and compute a Feynman diagram (or part of it) simply using the mouse and interacting with it by means of the tools available on the window. 
This panel can be divided in six groups of functions (as highlighted in Figure \ref{fig:panel}) presented in details in the following list. 

\begin{enumerate}[label=\arabic*)]
\item \textbf{Graphical environment}: In this portion of the interactive panel, the user can draw Feynman diagrams and act on them by means of the available tools interacting with the mouse. Any modification of the diagram is applied here in real time.
\item \textbf{Drawing tools}: These tools allow to draw/modify diagrams in the graphical environment.
\begin{itemize}
\item \textbf{Add vertices/propagators}:  Left- or Right-clicking in the graphical environment will place an isolated external (not integrated) point. Clicking on an already placed external point will turn it to an integrated one (black dot).  Any point is labeled by a number $k$ identifying its position $x_k$.
To add a scalar or fermionic propagator, the user has to left- or right-click respectively, then drag and release the mouse on a desired point of the interactive window. If the mouse is released in an empty spot, a new external point will be drawn in the corresponding position. Between two vertices it is allowed to draw multiple lines.
As soon as a propagator is placed, a pop-up window will appear. The user has to specify in the input field the weight of the propagator (numbers, letters and arithmetic operations) and press "\textit{OK}" to update it. 
\item \textbf{Remove/Move vertices}: Selecting those tools, the user can erase or move a vertex of the diagram and all the propagators attached to it. To erase a vertex it is sufficient to left-click on it while to move it the user has to left-click and drag it to the new desired position.
\end{itemize}
\item \textbf{Relation tools}: These tools allow the user to identify unique stars or triangles in the graph imposing the uniqueness relation $\sum_k\alpha_k=D/2$ with $k=1,2,3$ between the weights $\alpha_k$ as presented in section \ref{sec:startriangle}. If a star or a triangle is already unique by construction (when its weight is $D$ or $D/2$ respectively), these options are not needed. 
\begin{itemize}
\item \textbf{Add star/triangle relations}: 
In order to select the desired star directly on the graph, after selecting \texttt{Add star relations}, the user has to click on the black (integrated) vertex in which three propagators merge. To identify a triangle instead, after selecting \texttt{Add triangle relations}, one has to click on the three dots at the vertices of it.  In the case in which the same set of vertices identifies more than one triangle, a pop-up window containing a list of buttons with all the possible sub-triangle will appear. The desired sub-triangle can be selected simply clicking on the corresponding button.
\item \textbf{Clear selection/relations}: Clicking on these buttons, the user can deselect the vertices previously selected or reset the \texttt{STRrelations} function respectively.
\end{itemize} 
\item \textbf{Computation tools}: 
They include the relations of the uniqueness method of section \ref{sec:startriangle}.
\begin{itemize}
\item \textbf{Flip arrow}:
The user can flip the sign of a chosen fermionic propagator left- or right-clicking on the two vertices at the endpoints of it. 
\item \textbf{Merge}:
Selecting two vertices connected by more than one propagator, the user can combine them in a single one following the rules give in \eqref{merge1}, \eqref{merge2} and \eqref{merge3}.
\item \textbf{Chain rule}:
Clicking on an internal vertex connected to two propagators, the user can solve the integration by means of the chain rules \eqref{chain1}, \eqref{chain2} and \eqref{chain3}. 
\item \textbf{Star-triangle and Triangle-star}:
The user can select a star (triangle) in the diagram and solve (add) the integration on the internal vertex turning it into a triangle (star) according to the star-triangle relations \eqref{STRbos} and \eqref{STRferm}. 
The procedure to select stars and triangle is exactly the same presented for the tools \texttt{Add star relations} and \texttt{Add triangle relations}. 
\item \textbf{Clear prefactor}:
It resets the function \texttt{STRprefactor} at its initial value 1.
\end{itemize}
\item \textbf{Edit buttons}: 
The buttons \texttt{Undo} and \texttt{Redo} cancel and restore the most recent action respectively. The \texttt{Clear All} button resets the working-space and all the functions to their initial values.
\item \textbf{Output tools}: 
The \texttt{Print} buttons show the content of the functions \texttt{STRrelations}, \texttt{STRprefactor} and \texttt{STRintegral} below the \texttt{STR} interactive panel. With the  \texttt{Export} menu it is possible to export to the Mathematica kernel the data in the functions \texttt{STRprefactor}, \texttt{STRintegral} and \texttt{STRrelations} and the diagram shown in the graphical environment as a \texttt{Graphics[]} object in \texttt{STRgraph}.
\end{enumerate}
In order to load the data generated by the function \texttt{STR} in the Mathematica kernel, one has to use the \texttt{Export} button. For instance, once the user draws the desired diagram, it is possible to export the graph itself and its integral representation (\texttt{STRprefactor}$\int$\texttt{STRintegral}) as follows 
\begin{equation*}
\vcenter{\hbox{\includegraphics[trim={1.2cm 0 0 0},clip,width=4cm]{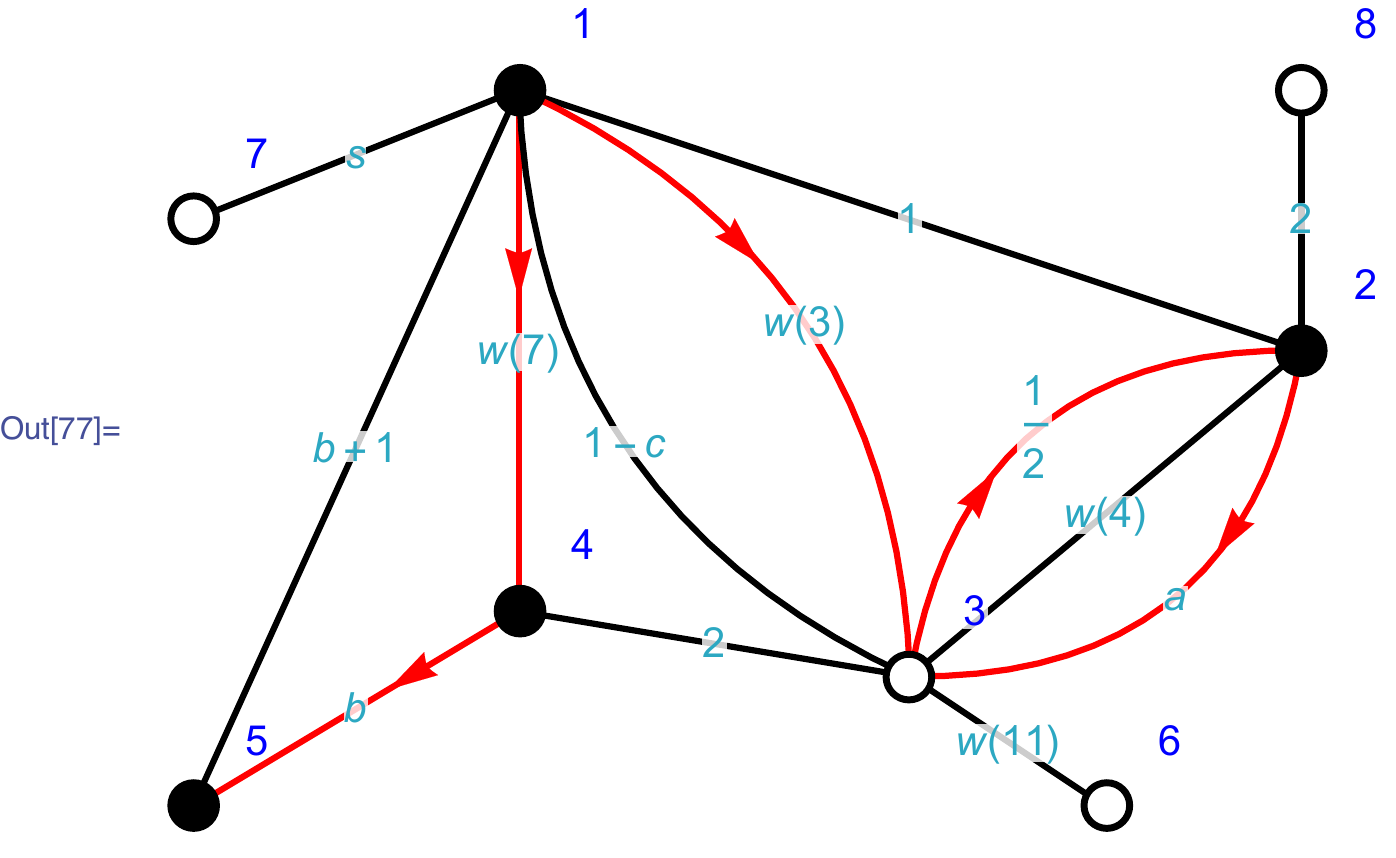}}}\quad\vcenter{
\hbox{\includegraphics[trim={1.5cm 0 0 0},clip,width=2.5cm]{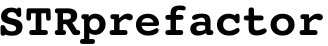}}\hbox{\includegraphics[trim={1.4cm 0 0 0 },clip,width=.3cm]{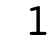}}\hbox{\includegraphics[trim={1.5cm 0 0 0},clip,width=2.5cm]{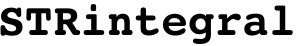}}\;\;\hbox{\includegraphics[trim={1.4cm 0 0 0},clip,width=11cm]{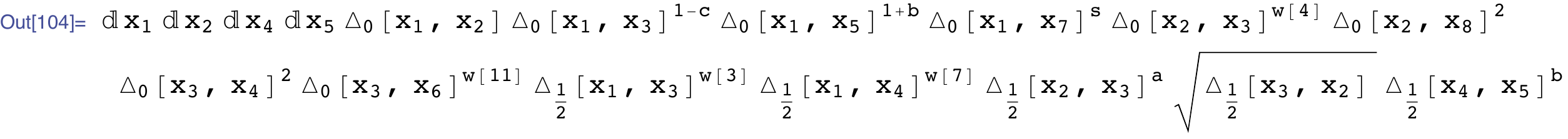}}}
\end{equation*}
where the symbol $\vcenter{\hbox{\includegraphics[trim={1.5cm 0 0 0},clip,width=.8cm]{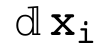}}}\Rightarrow d^Dx_i$ is the integration measure and $\vcenter{\hbox{\includegraphics[trim={1.6cm 0 0 0},clip,width=2cm]{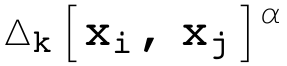}}}\Rightarrow\frac{{\slashed{x}_{ij}}^{2k}}{(x_{ij}^2)^{\alpha+k}}$ the propagators \eqref{prop}.
At this stage it is possibe for instance to impose the uniqueness conditions to a triangle and a star using the \texttt{Relation tools} and export them updating the function \texttt{STRrelations}.
A possible choice is the following ($D=d$)
\vspace*{6px}

$
\vcenter{\hbox{\includegraphics[trim={1.5cm 0 0 0},clip,width=2.5cm]{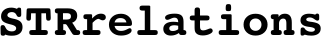}}\hbox{\includegraphics[trim={1.4cm 0 0 0 },clip,width=6.5cm]{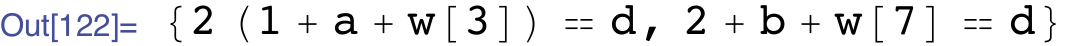}}}
$
\vspace*{6px}

\noindent
where we identified the star in $x_4$ and a Yukawa triangle with vertices $x_1$, $x_2$ and $x_3$ as unique.
Then, using the \texttt{Computation tools}, one can solve part the diagram as follows
\begin{equation*}
\vcenter{\hbox{\includegraphics[trim={1.2cm 0 0 0},clip,width=3.5cm]{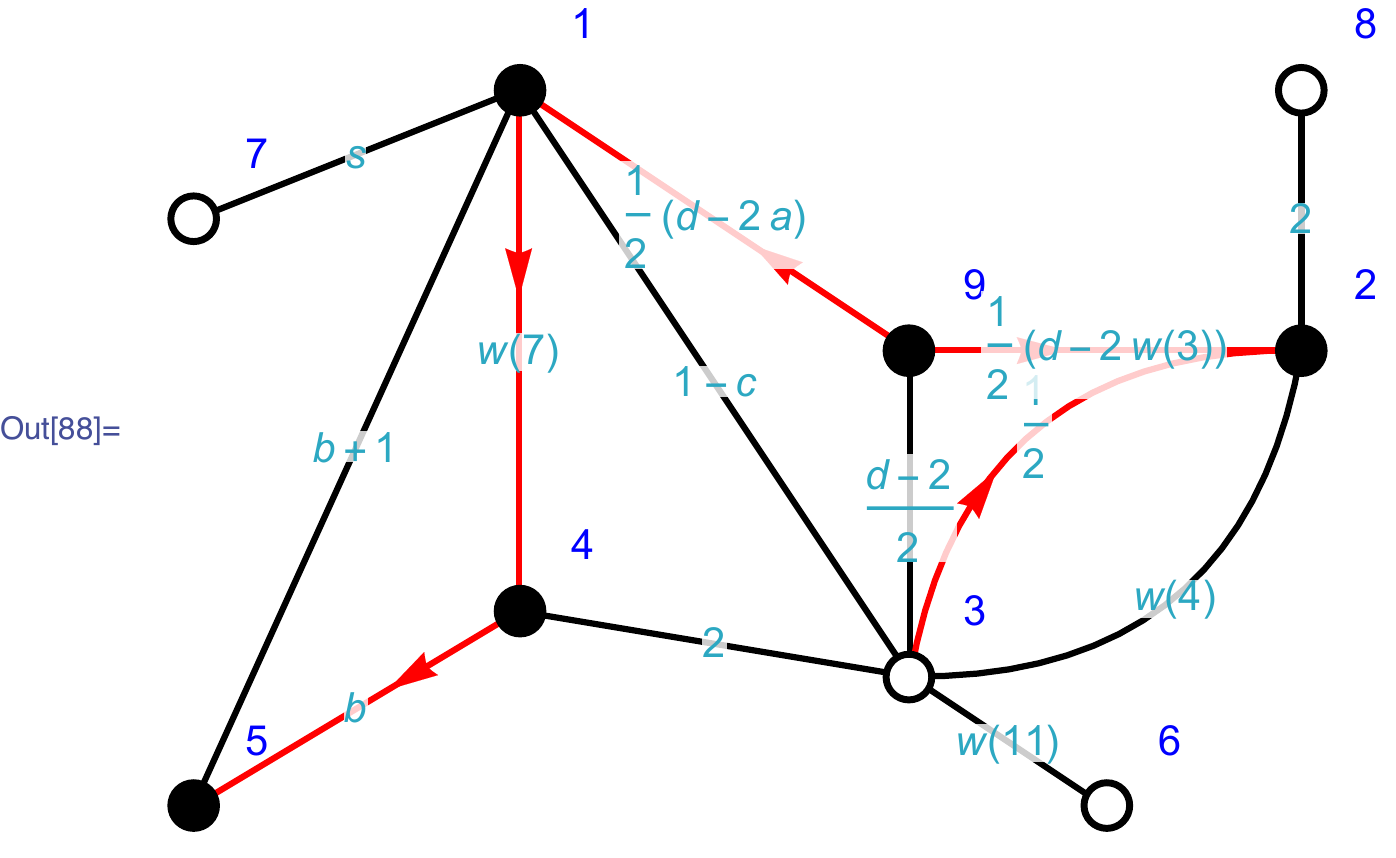}}}\Rightarrow
\vcenter{\hbox{\includegraphics[trim={1.2cm 0 0 0},clip,width=3.5cm]{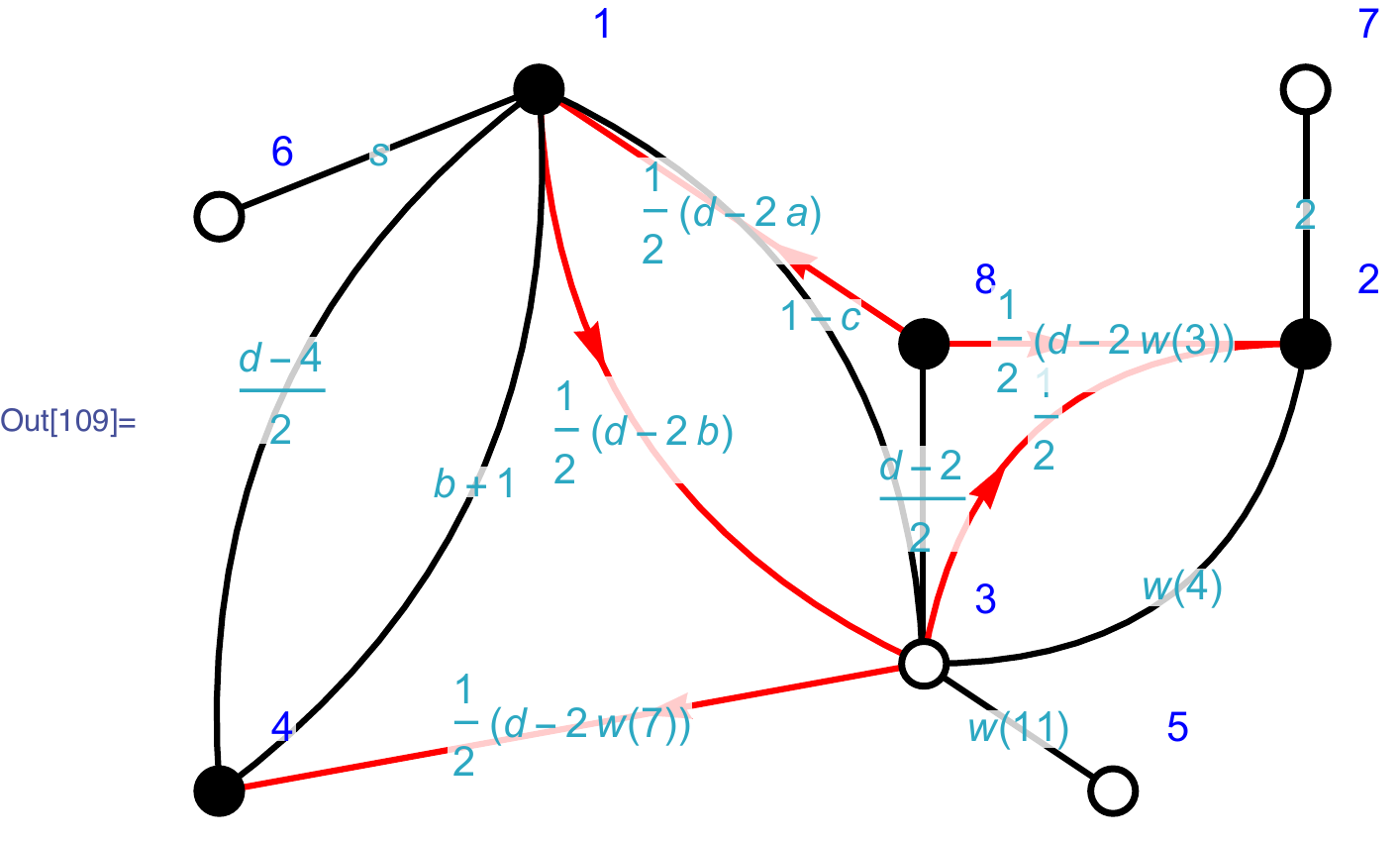}}}\Rightarrow
\vcenter{\hbox{\includegraphics[trim={1.2cm 0 0 0},clip,width=3.5cm]{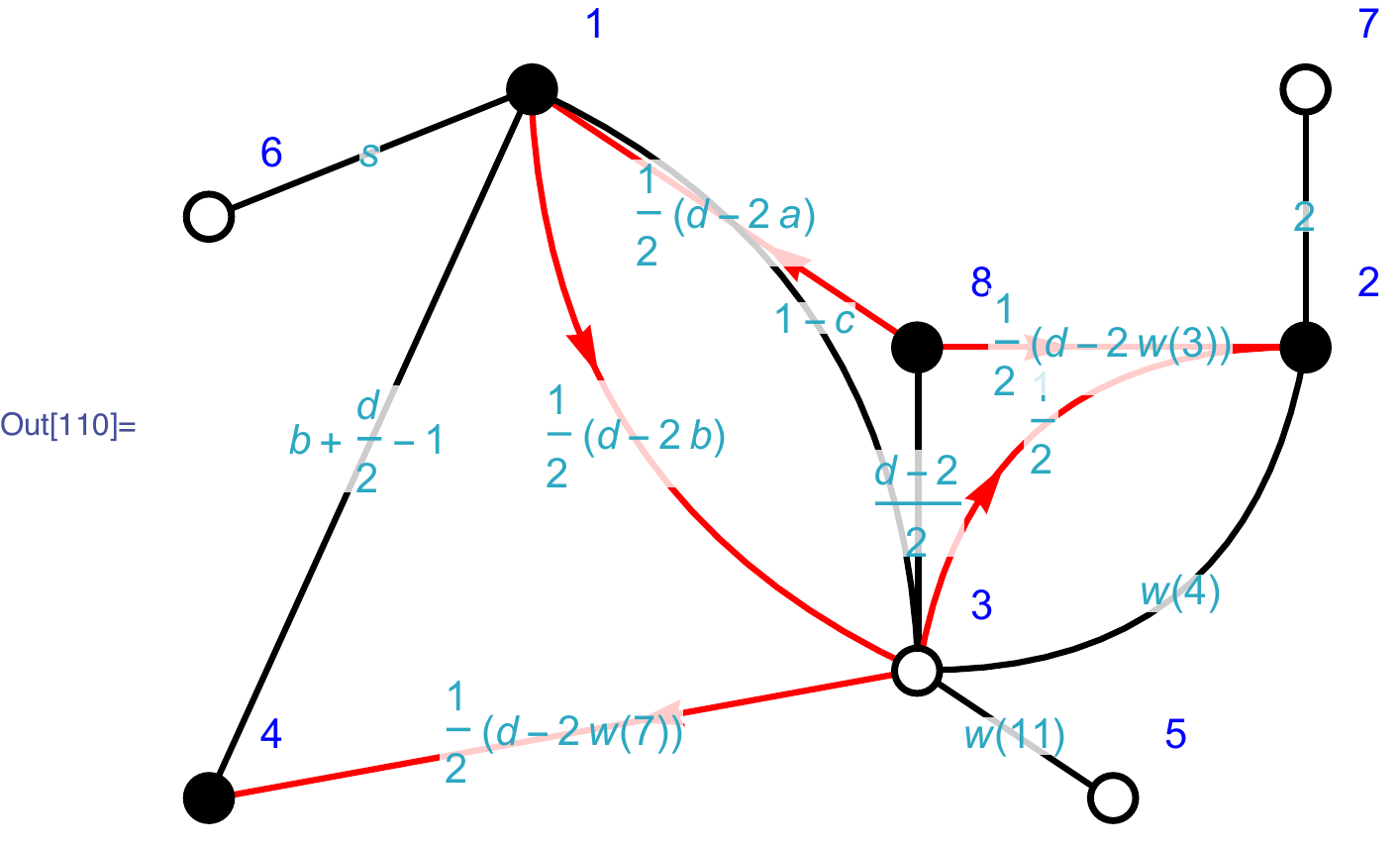}}}\Rightarrow
\vcenter{\hbox{\includegraphics[trim={1.2cm 0 0 0},clip,width=3.5cm]{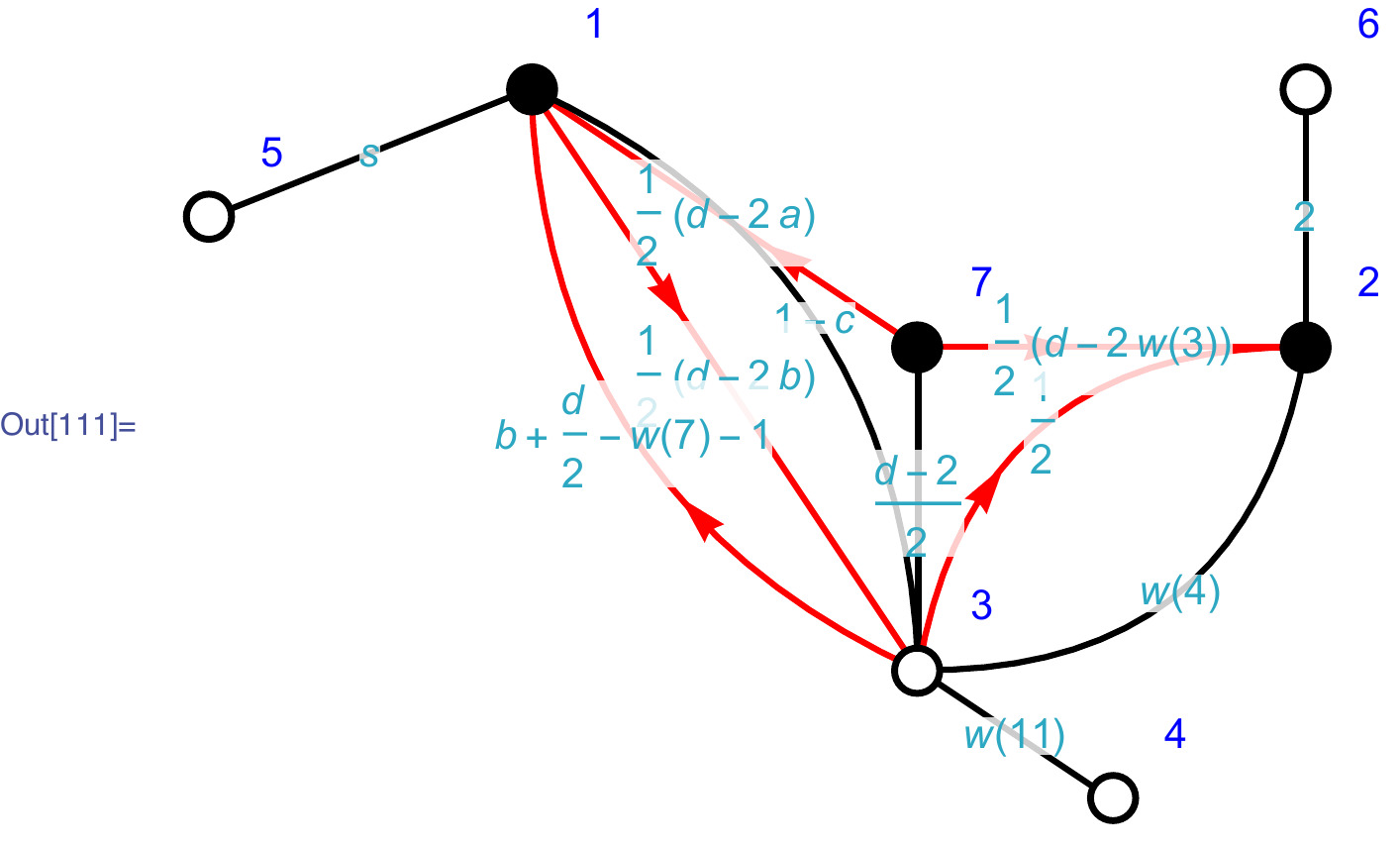}}}
\end{equation*}
Acting with the method of uniqueness, we reduced the number of integrations from five to three. The functions generated in the process are stored in  \texttt{STRprefactor}, then exporting the data at the end of the computation we have
\begin{equation*}
\vcenter{
\hbox{\includegraphics[trim={1.5cm 0 0 0},clip,width=2.5cm]{STRprefactor}}\hbox{\includegraphics[trim={1.4cm 0 0 0 },clip,width=12cm]{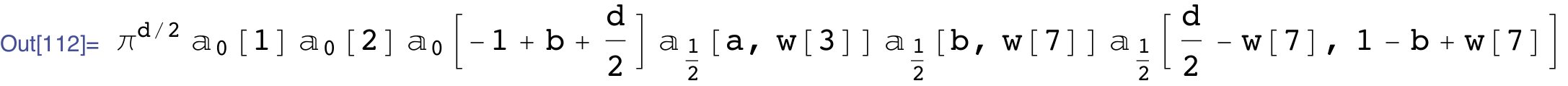}}\hbox{\includegraphics[trim={1.5cm 0 0 0},clip,width=2.5cm]{STRintegral}}\hbox{\includegraphics[trim={1.4cm 0 0 0},clip,width=15cm]{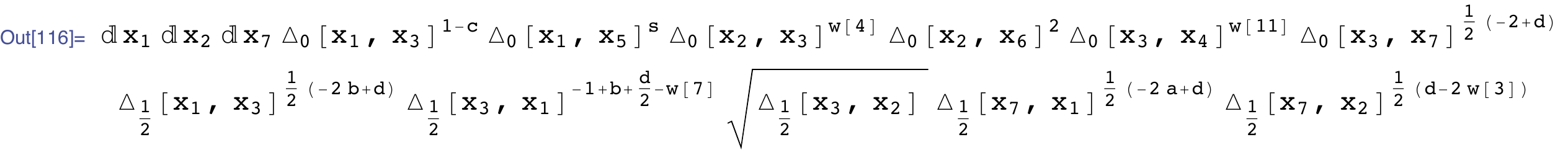}}}
\end{equation*}
where $\vcenter{\hbox{\includegraphics[trim={1.5cm 0 0 0},clip,width=4cm]{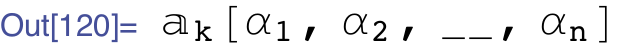}}}\;\Rightarrow\; \mathbb{a}_{k}(\alpha_1,\alpha_2,...,\alpha_n)$
defined in \eqref{defa} and \eqref{amultiple}.

\section{The package at work}
\label{sec:example}

Here we present an example of computation of a Feynman diagram involving Yukawa star-triangle relations in $D=4$. 
This graph arises in the computation of the anomalous dimension of length-2 operators \cite{Kazakov:2018gcy} in the $\gamma$-deformed $\mathcal{N}=4$ SYM theory in a double scaling limit\footnote{Similar double scaling limits were also studied in \cite{Correa:2012nk,Bonini:2016fnc,deLeeuw:2016vgp,Aguilera-Damia:2016bqv,Preti:2017fhw} in different contexts.} that combine large imaginary twists and the weak coupling limit \cite{Gurdogan:2015csr}. 

Running the command \texttt{STR[4]} and drawing the diagram (the first one in the following), one can fully solve it through a specific sequence of applications of \texttt{Computation tools}, namely
\begin{equation*}\begin{split}
&\vcenter{\hbox{\includegraphics[trim={1.75cm 0 0 0},clip,width=3cm]{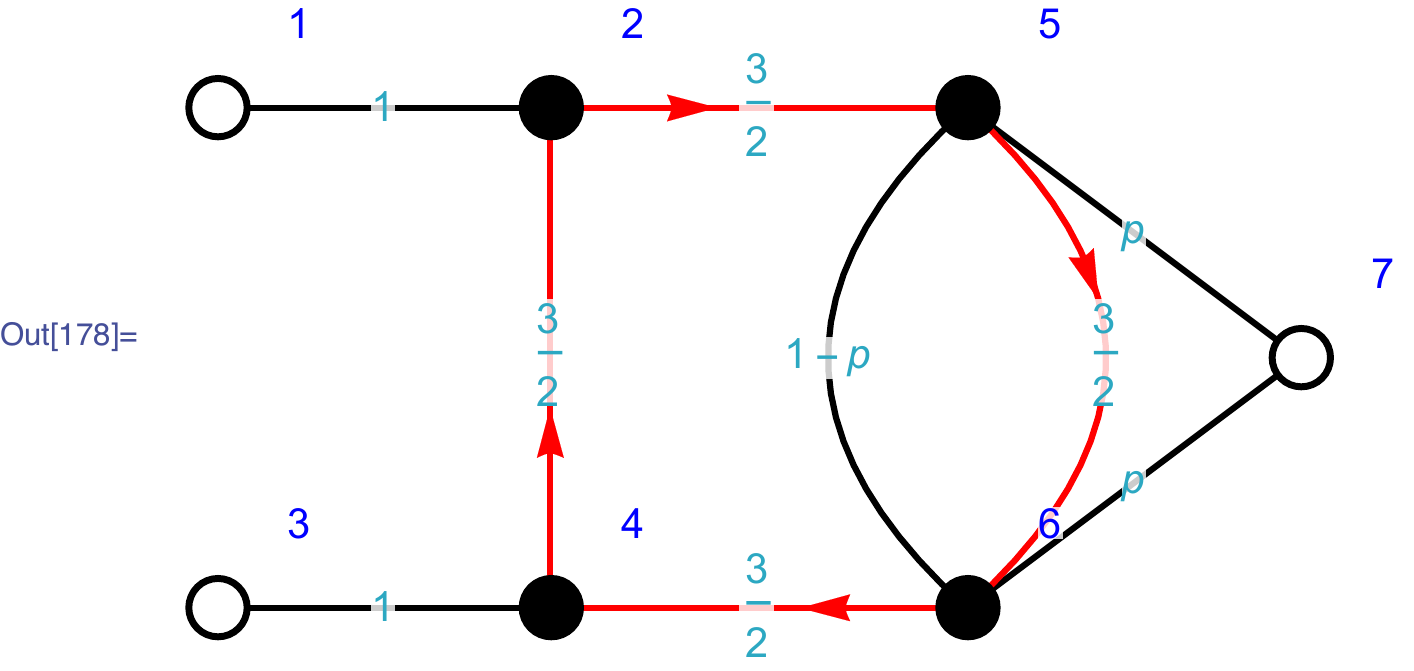}}}\!\!\Rightarrow\!\!
\vcenter{\hbox{\includegraphics[trim={1.75cm 0 0 0},clip,width=3cm]{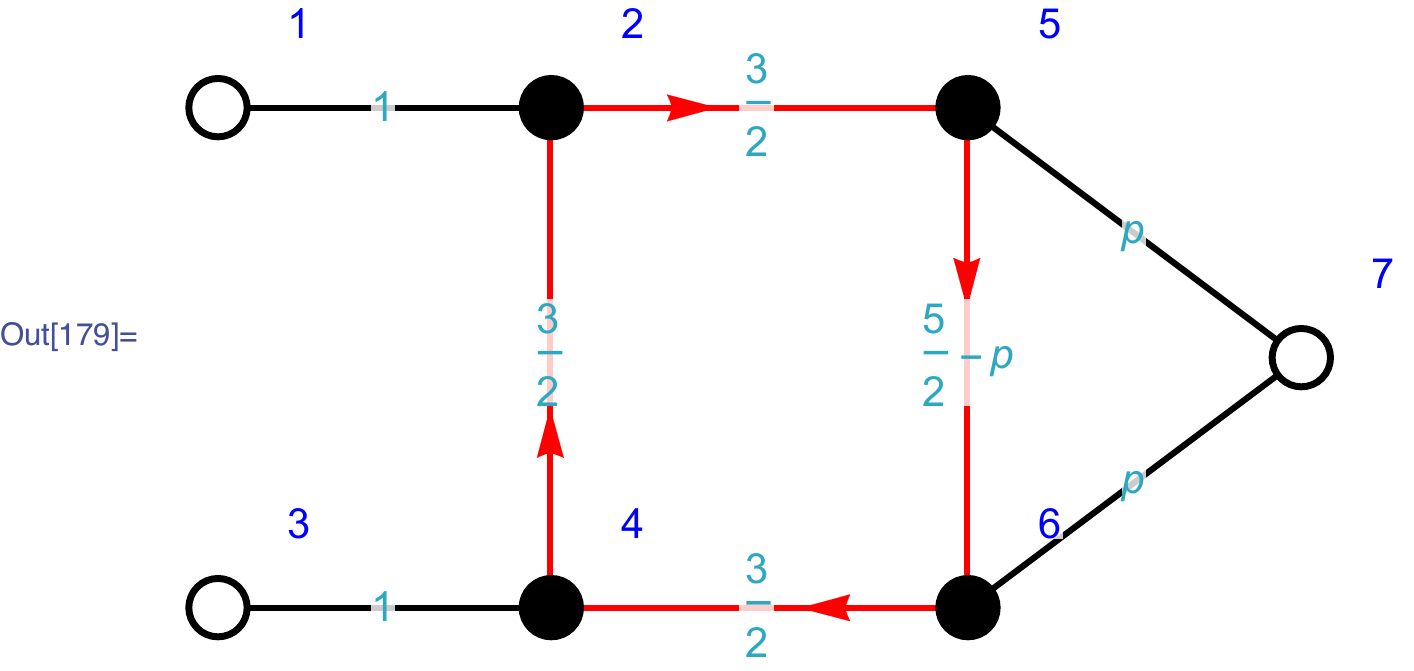}}}\!\!\Rightarrow\!\!
\vcenter{\hbox{\includegraphics[trim={1.75cm 0 0 0},clip,width=3cm]{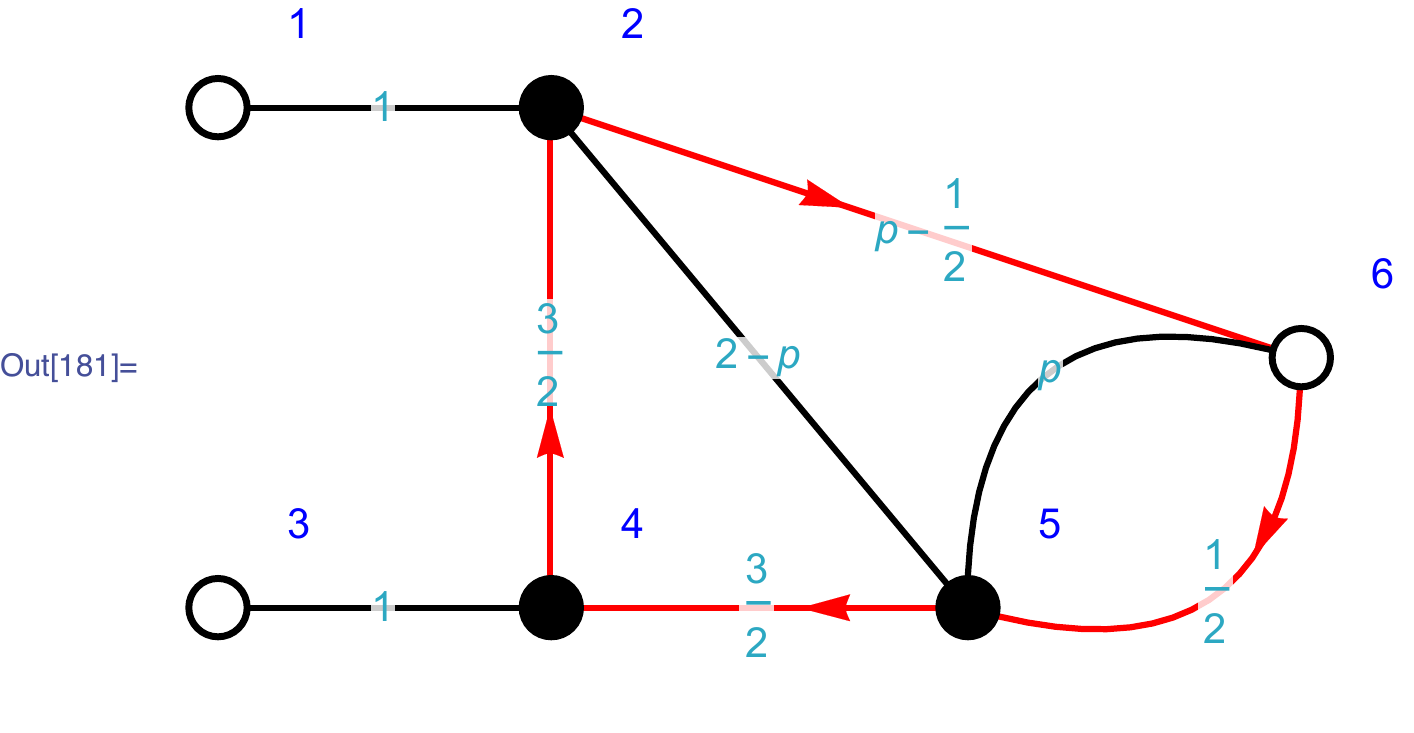}}}\!\!\Rightarrow\!\!
\vcenter{\hbox{\includegraphics[trim={1.75cm 0 0 0},clip,width=3cm]{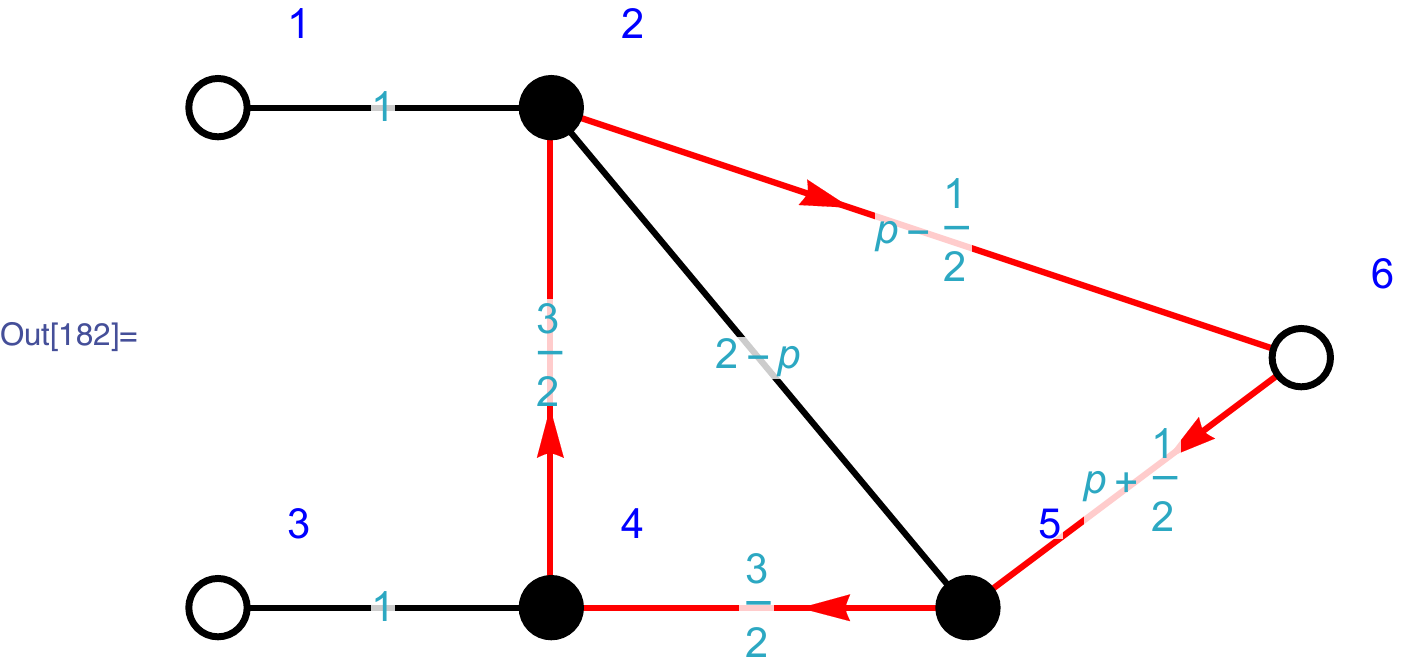}}}\!\!\Rightarrow\!\!\\
\!\!\Rightarrow&
\vcenter{\hbox{\includegraphics[trim={1.75cm 0 0 0},clip,width=3cm]{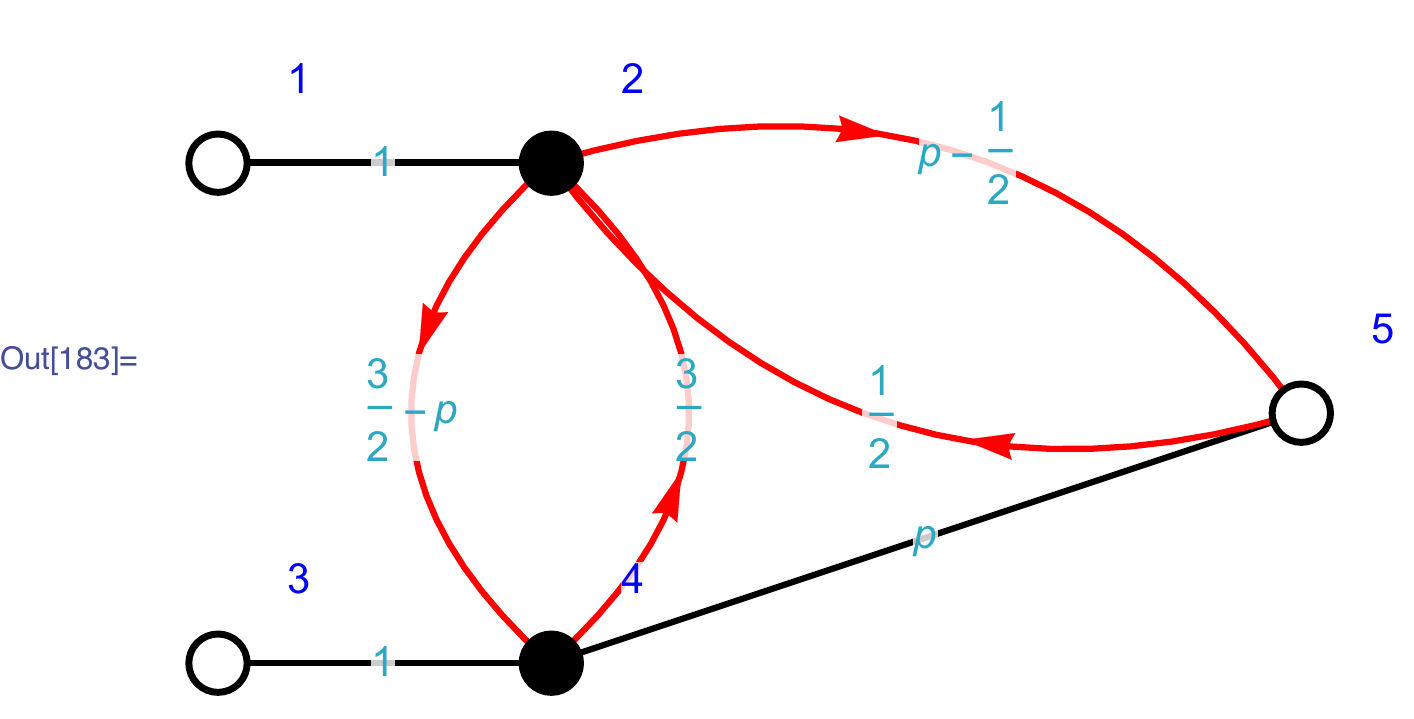}}}\!\!\Rightarrow\!\!
\vcenter{\hbox{\includegraphics[trim={1.75cm 0 0 0},clip,width=2.2cm]{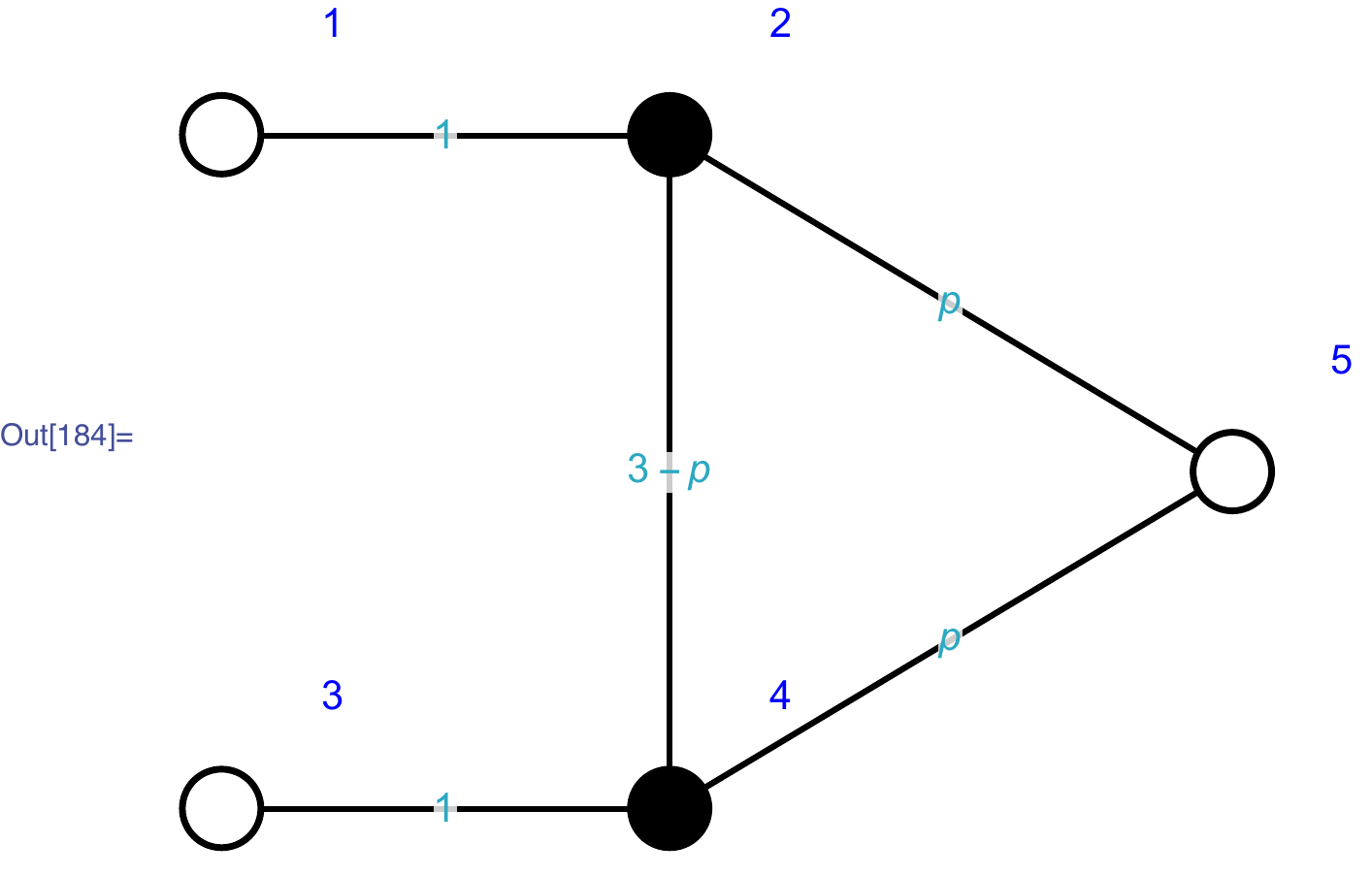}}}\!\!\Rightarrow\!\!
\vcenter{\hbox{\includegraphics[trim={1.75cm 0 0 0},clip,width=2.2cm]{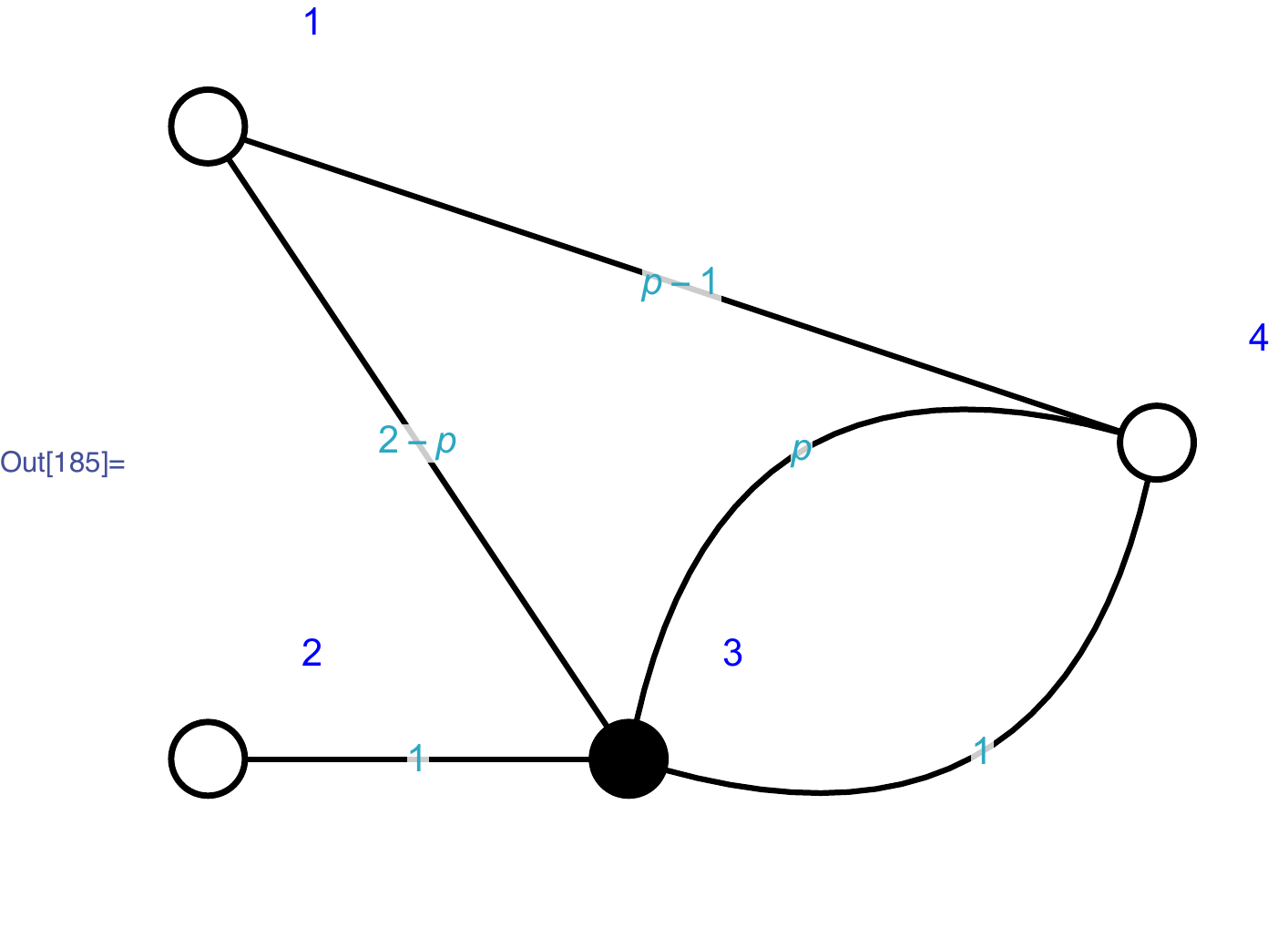}}}\!\!\Rightarrow\!\!
\vcenter{\hbox{\includegraphics[trim={1.75cm 0 0 0},clip,width=2.2cm]{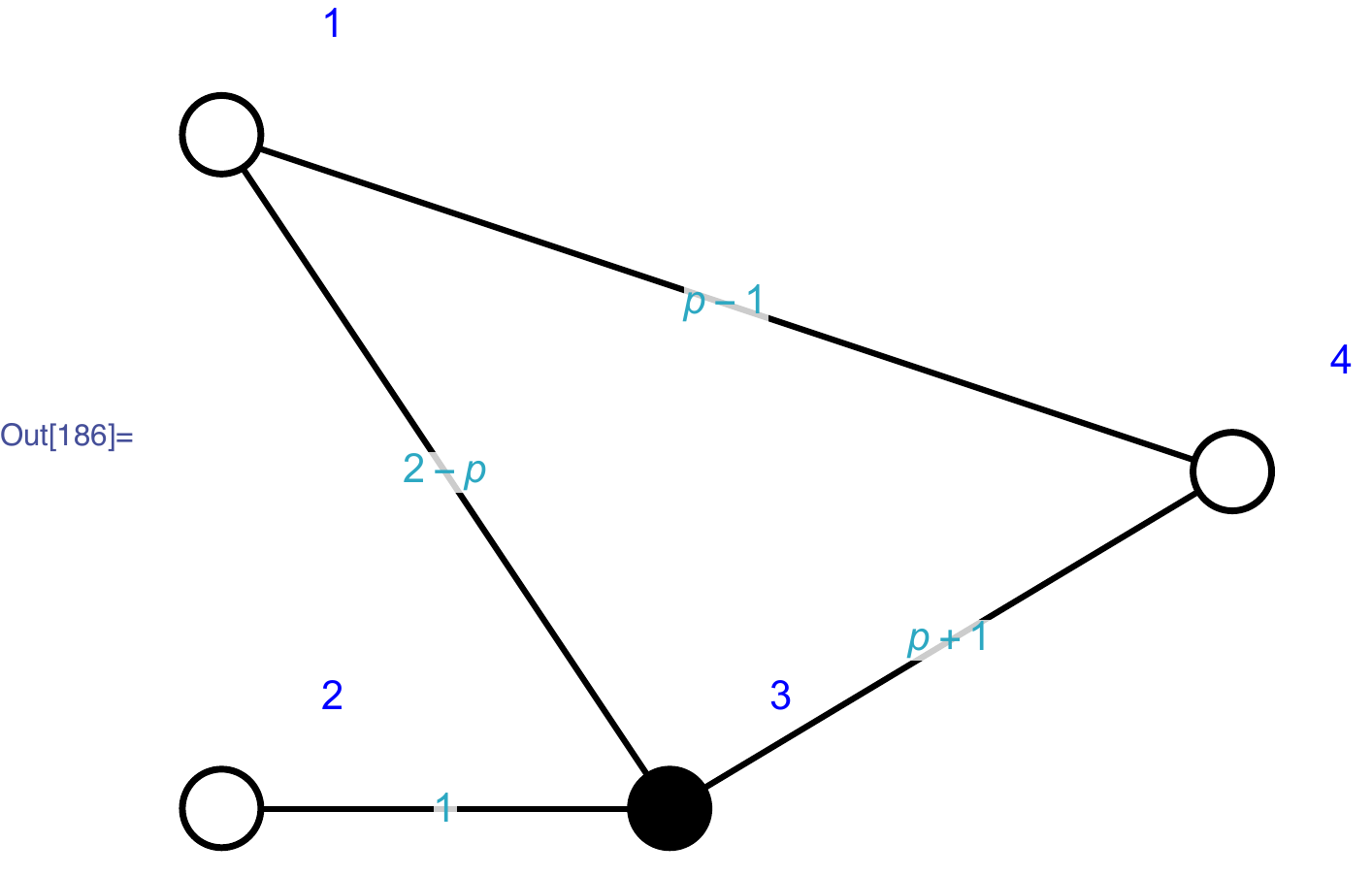}}}\!\!\Rightarrow\!\!
\vcenter{\hbox{\includegraphics[trim={1.75cm 0 0 0},clip,width=2.2cm]{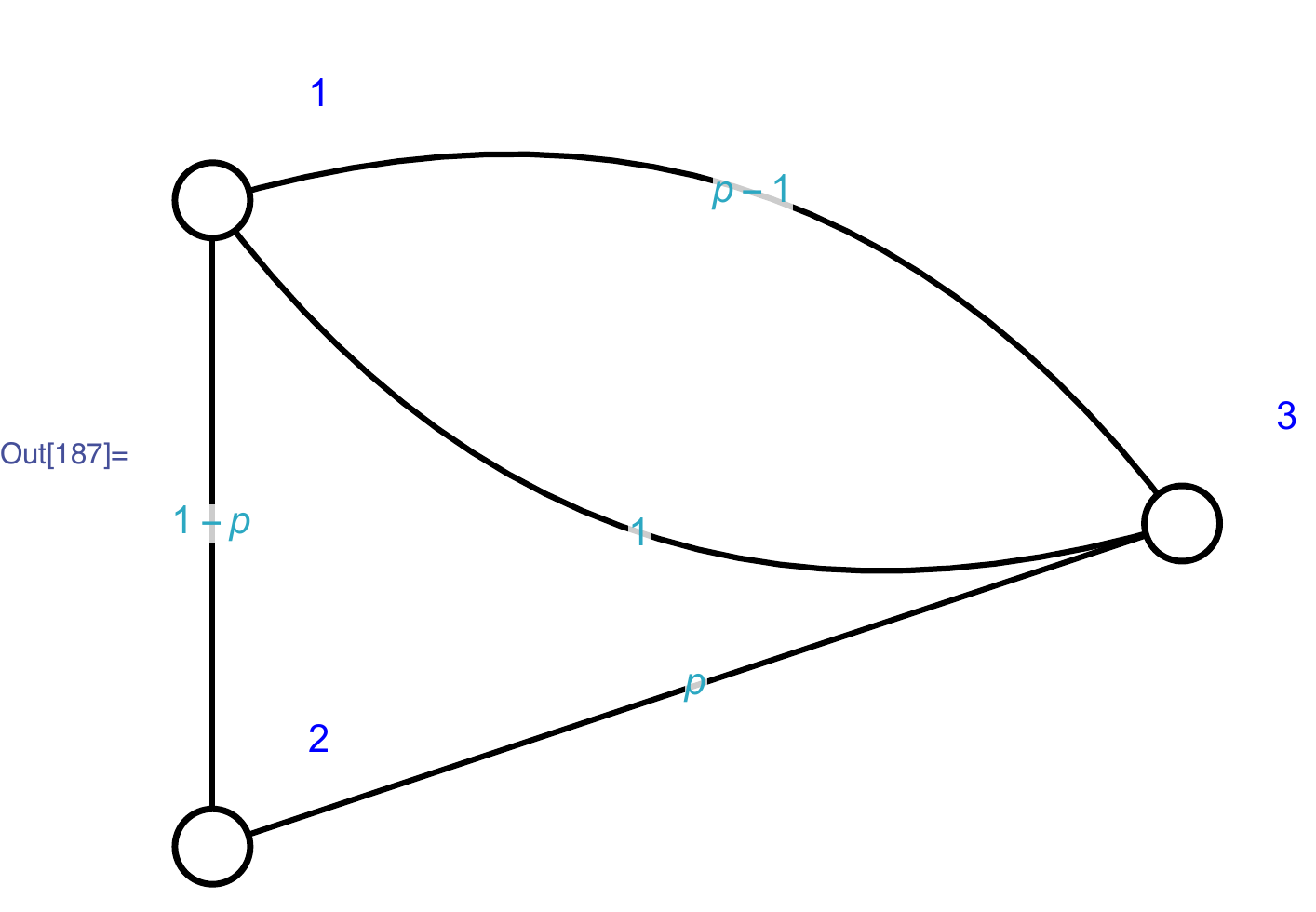}}}\!\!\Rightarrow\!\!
\vcenter{\hbox{\includegraphics[trim={1.75cm 0 0 0},clip,width=1.5cm]{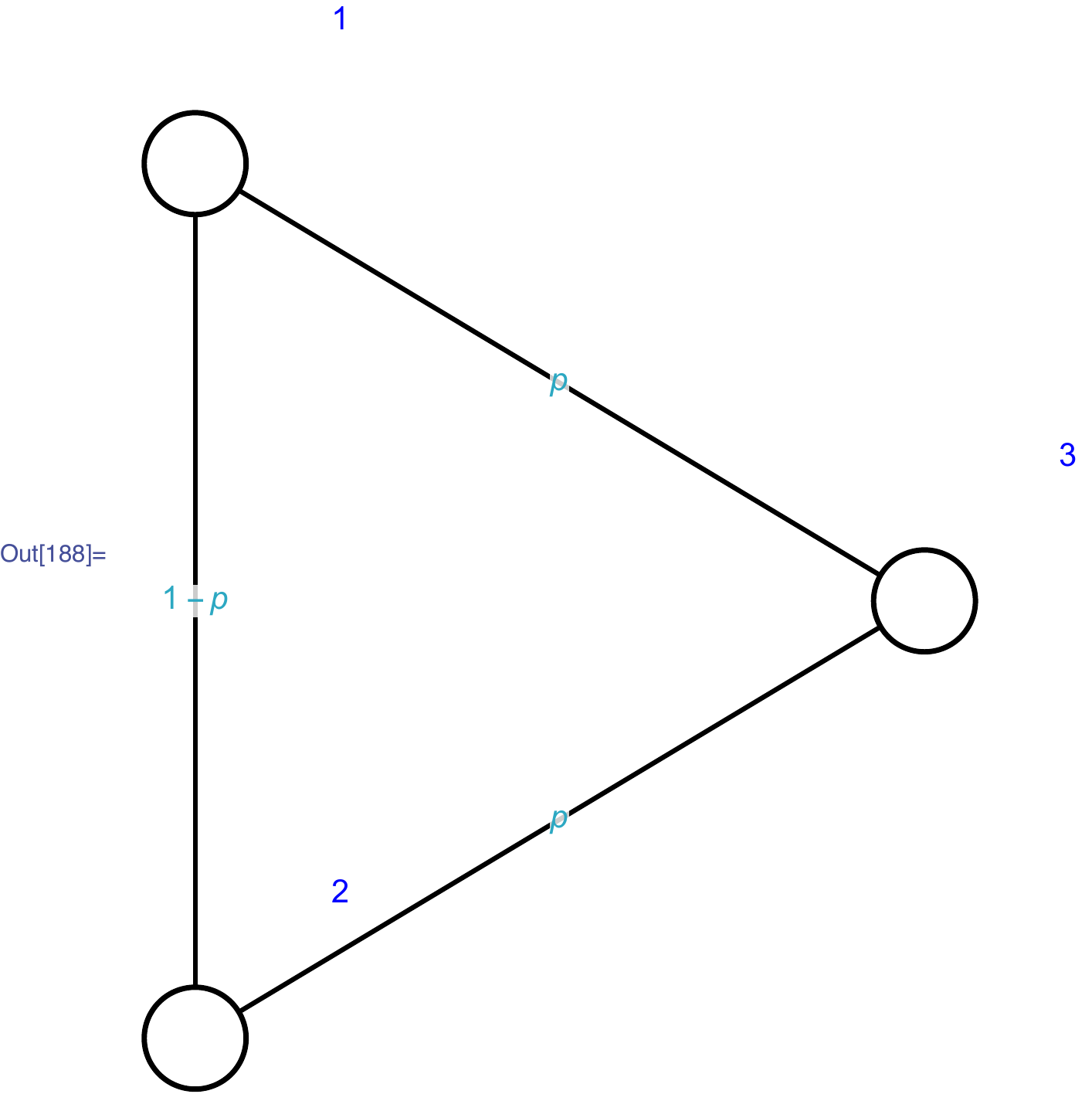}}}
\end{split}\end{equation*}
where we used the tools \texttt{Star-triangle} and \texttt{Merge} several times. In the last step no residual integrations remain, then exporting the data the output functions are
\vspace*{10px}

$
\vcenter{
\hbox{\includegraphics[trim={1.5cm 0 0 0},clip,width=2.5cm]{STRprefactor}}\hbox{\includegraphics[trim={1.4cm 0 0 0 },clip,width=12cm]{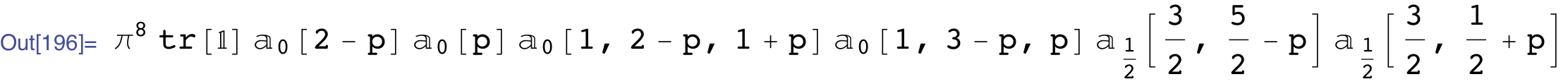}}\hbox{\includegraphics[trim={1.5cm 0 0 0},clip,width=2.5cm]{STRintegral}}\hbox{\includegraphics[trim={1.4cm 0 0 0},clip,width=7cm]{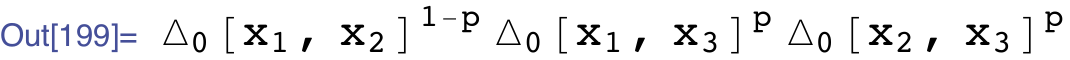}}}
$
\vspace*{10px}

\noindent 
Finally,  using the function \texttt{STRSimplify}, we can conclude that
\begin{equation*}
\vcenter{\hbox{\includegraphics[trim={1.55cm 0 0 0},clip,width=3cm]{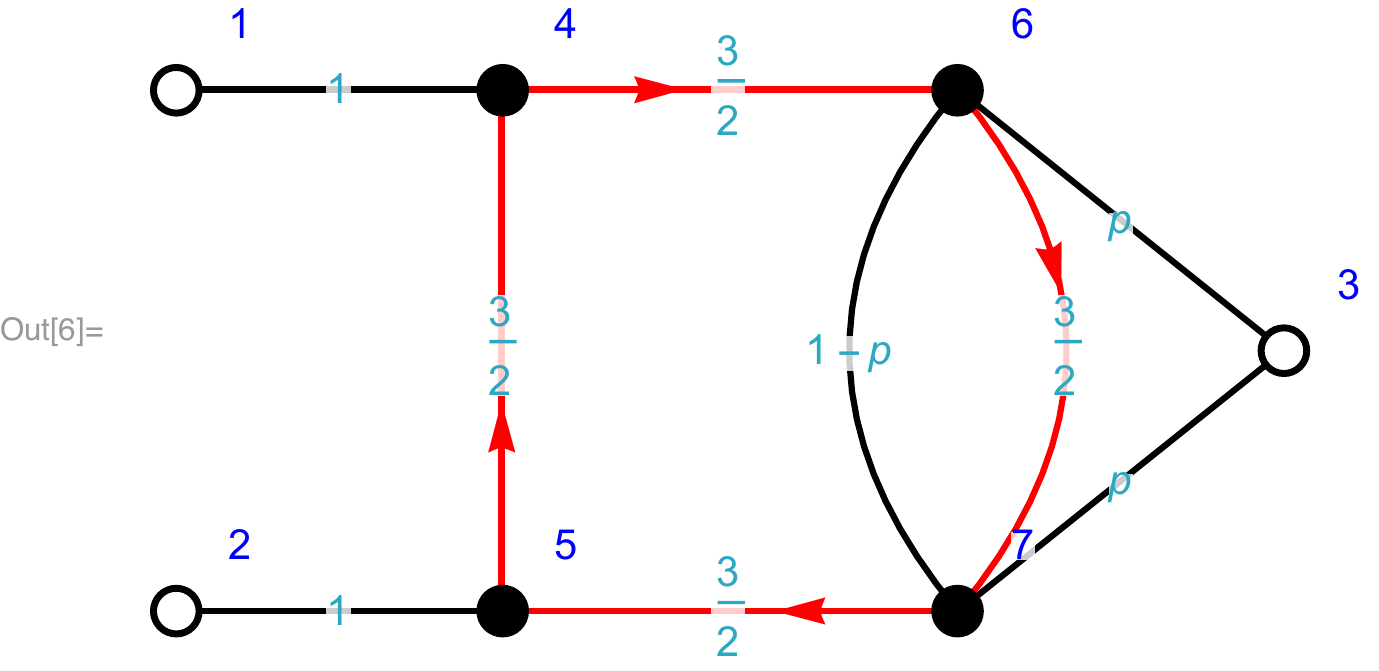}}}=
\vcenter{
\hbox{\includegraphics[trim={1.5cm 0 0 0},clip,width=12cm]{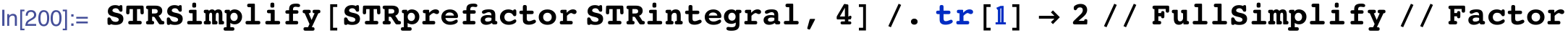}}\hbox{\includegraphics[trim={1.4cm 0 0 0 },clip,width=8cm]{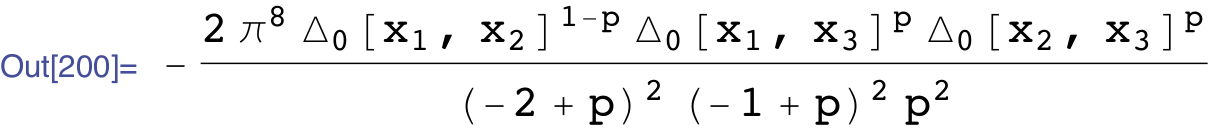}}}
\end{equation*}
where we used that $\text{tr}(\mathbb{1})=2$ in $D=4$.

\section*{Acknowledgments}
We thank Andrei Kataev for his invitation to present this work at ACAT 2019.

\section*{References}
\bibliographystyle{iopart-num}

\bibliography{biblio}

\end{document}